\documentclass{article}
\usepackage{graphicx} % Required for inserting images

\usepackage{amsmath,amssymb}
\usepackage{hyperref}
\usepackage{subcaption}
\usepackage{bbm}
\usepackage[numbers,compress]{natbib}
\usepackage[margin=1.2in]{geometry}

\input{preamble}

\newcommand{\der}[0]{\mathrm{d}}

\newcommand{\was}[0]{\mathcal{W}_2}

\newcommand{\sca}[2]{\langle #1,#2 \rangle}
\newcommand{\ind}[0]{\mathbbm{1}}

\title{Minimally dissipative multi-bit logical operations}

\author{J\'{e}r\'{e}mie Klinger and Grant M. Rotskoff}
\date{\today}

\begin{document}

\maketitle

\begin{abstract}
Modern computing architectures are vastly more energy-dissipative than fundamental thermodynamic limits suggest, motivating the search for principled approaches to low-dissipation logical operations. 
We formulate multi-bit logical gates (bit erasure, NAND) as optimal transport problems, extending beyond classical one-dimensional bit erasure to scenarios where existing methods fail.
Using entropically regularized unbalanced optimal transport, we derive tractable solutions and establish general energy-speed-accuracy trade-offs that demonstrate that faster, more accurate operations necessarily dissipate more energy. 
Furthermore, we demonstrate that the Landauer limits cannot be trivially overcome in higher dimensional geometries.
We develop practical algorithms combining optimal transport with generative modeling techniques to construct dynamical controllers that follow Wasserstein geodesics. 
These protocols achieve near-optimal dissipation and can, in principle, be implemented in realistic experimentally set-ups. 
The framework bridges fundamental thermodynamic limits with scalable computational design for energy-efficient information processing.
\end{abstract}

\paragraph{Introduction}

The high energetic cost of modern computing architectures has motivated a renewed interest in both low-dissipation computing paradigms and also the fundamental, physical limits of efficiency imposed by the Second Law of Thermodynamics.
Despite the importance of this problem, generic design principles for a complete set of practical and efficient logical operations have not yet been articulated. 
This problem has a canonical limit, due to Landauer \cite{landauer_irreversibility_1961}, who showed that the process of quasistatically resetting or \emph{erasing} a bit of information incurs at least an energetic cost of $k_{\rm B} T \log 2$, but, of course, requires an infinitely slow protocol.
The decades of optimization of processor design, however, have largely focused on performance over thermodynamic efficiency and modern computers are vastly more dissipative than the thermodynamic optimum.
The intuitive trade-off between speed and dissipation suggests optimizing on the Pareto front of these objectives, an approach that has been pursued theoretically~\cite{aurell_optimal_2011,Bechhoefer2020} using tools from stochastic thermodynamics~\cite{sekimoto_langevin_1998}.
Recent results in the nonequilibrium statistical mechanics literature that better articulate the trade-offs among energy, speed, and accuracy~\cite{Saito2023, KR25_1} create the opportunity to better understand the finite-time limits of thermodynamic computing.
Coupling these tools with computational advances in optimal transport, generative modeling, and stochastic optimal control simultaneously enables us to provide explicit, experimentally testable predictions for a Pareto optimal logical operations. 

The precise connection between optimal transport distances and dissipation~\cite{benamou_computational_2000}, which Aurell extended to stochastic thermodynamics~\cite{aurell_optimal_2011,aurell_refined_2012}, is leading to improved, scalable algorithms for nonequilibrium protocol design~\cite{KR25_1}.
Here, we leverage this connection to implement minimally dissipative logical operations as solutions to optimal transport minimization problems, enabling detail-agnostic energetic bounds.
We do so for all core logical gates---OR, AND, NAND---which typically operate on multiple bits, a set-up in which all previous existing methods fail.
To carry out this task, we extend recent work from the computational optimal transport literature~\cite{Chizat2018}, and generative modeling techniques \cite{hyvarinen_estimation_2005,song_score-based_2022}, we subsequently propose an algorithmic route toward designing per-operation optimal controllers.
To carry out this task, we extend recent work from the computational optimal transport literature~\cite{Chizat2018}, and subsequently propose an algorithmic route based on generative modeling techniques \cite{hyvarinen_estimation_2005,song_score-based_2022} toward designing per-operation optimal controllers.
The resulting space- and time-dependent controllers are directly implementable using virtual potentials, for example real time feedback optical traps \cite{jun_high-precision_2014}, paving the way for experimental exploration of these optimal logical operations.

The topic of optimal control for logical operation has been studied with a variety of model systems, but the paradigmatic double-well thought-experiment of single bit erasure introduced by Landauer \cite{landauer_irreversibility_1961} remains the basis for much of the theoretical and experimental developments on this topic.
In this set-up, the state of a bit is encoded by the position of a particle relative to some fixed reference frame; for example, in one-dimension, $x>0$ encodes state 1,and $x\leq 0$ encodes state 0.
We assume a finite-temperature Boltzmann distribution for the particle with an energy function that ensures that each state is metastable, the simplest choice is a conventional double-well potential with the barrier centered at $x=0$.
The \emph{bit erasure} problem can then be formulated in terms of the probability distribution of the particle: erasure simply means that all probability mass is evacuated from one state, for example, by modulating the potential energy in a time-dependent fashion. 
In the quasistatic driving limit, approaches combining linear-response theory \cite{DeWeese2014} and thermodynamics geometry \cite{sivak_thermodynamic_2012} have led to explicit protocol design for various confining potentials.
In parallel, improved techniques for sculpting potential energy surfaces of optically trapped colloidal particles have enabled increasingly precise experimental tests of the Landauer limit in the quasistatic regime \cite{Dillenschneider2009, berut_experimental_2012,Bechhoefer2016}.
For faster driving, alternative counterdiabatic parametric protocols \cite{boyd_shortcuts_2022} yield precise but suboptimal finite-time bit erasure.

The framework for optimal control based on linear response and thermodynamic geometry assumes a set of control parameters that are linearly coupled to observables of the controlled system.
This assumption reflects experimental constraints on which degrees of freedom can be controlled, but also explicitly limits controllability.
Expanding the control problem to the full probability distribution of the system, while increasing protocol complexity, reframes the finite time optimization problem without the assumption of linear response.
Determining the minimally dissipative protocol hence becomes a problem of transforming a given probability distribution into a fixed target, which is known as \emph{optimal transport}. 

Classically, the optimal transport problem seeks a sequence of distributions that interpolate between an initial and target distribution while minimizing some cost functional~\cite{villani_topics_2003}.
The connection between optimal transport distances and dissipative gradient flows has its origins in the applied mathematics literature: the groundbreaking work of \cite{jordan_variational_1998} established that the relaxation of the Fokker-Planck equation corresponds to a geodesic measured by the optimal transport distance. 
Benamou and Brenier subsequently showed that the advection equation from fluid mechanics provided a variational principle for computing optimal transport distances, which Aurell subsequently connected to stochastic thermodynamics~\cite{aurell_refined_2012}.
The observation that minimizing these distances between probability distributions was equivalent to optimal, finite-time control has been utilized to obtain a direct solution to the classical Landauer double-well bit erasure problem~\cite{Bechhoefer2020}.
Because the basic objects in this control problem are probability distributions, the target state can be defined with additional flexibility, for example, leaving a small, but non-zero amount of probability in one of the wells after a finite time~\cite{Bechhoefer2020_2}. 
For strictly one-dimensional systems, the resulting protocols can be determined exactly and have recently been experimentally realized in single particle driving experiments \cite{oikawa2025experimentallyachievingminimaldissipation}.
Similar connections have been explored in the finite state space thermodynamical framework of Markov jump process \cite{Saito2023}, yielding theoretical bit erasure bounds \cite{Park2022}, but generic protocol design solutions have yet to be identified. Likewise, thermodynamics of quantum systems and computers are a blossoming research field~\cite{Deffner2022}.

While the abstract formulation of the optimal transport problem introduces the complexity of controlling the full distribution, a number of recent developments in computational optimal transport~\cite{peyre_computational_2019} and machine learning~\cite{tong_improving_2023} bring even complex, high-dimensional distributions into reach.
The Sinkhorn algorithm, a computationally efficient fixed point iteration, provides a highly scalable estimator for suitably regularized optimal transport distances.
We show that the logical gate operations can be reformulated as an unbalanced optimal transport problem and develop an efficient Sinkhorn-type algorithm to solve it. 
These solutions can subsequently be leveraged to build dynamical control protocols~\cite{chennakesavalu_unified_2023,KR25_1}, which we do here using the approach of measure transport generative modeling~\cite{liu_flow_2022, albergo_stochastic_2023, lipman_flow_2022}.
The combination of these disparate techniques provides an elegant and integrated solution to the problem of nonequilibrium protocol design for logical operations. 
A number of \emph{physical} questions about the nature of control emerge from this framework: i) how does dimensionality affect the limits established by Landauer? ii) how big is the cost of finite time driving in comparison to the linear response limit? iii) how do inaccuracies in the protocol optimization impact protocol performance?
Both our generic formulation of the control problem and the specific protocols we develop yield provide insight into these essential questions.

The paper is organized as follows. In section \ref{seq:theory} we define the core two-bit operations---full erasure, partial erasure \cite{Bechhoefer2020_2} and NAND gate \cite{Crutchfield2025}---in terms of two-dimensional single particle control problems.
Leveraging the optimal transport formulation, we then establish a number of related thermodynamic energy-speed-accuracy trade-offs \cite{Saito2023}.
In Section \ref{seq:regularized-lg}, we introduce an entropically regularized formulation \cite{cuturi_sinkhorn_2013} of the logical operations. Using the computational optimal transport tools \cite{Chizat2018}, we analyze the energy - regularization dependence, and propose an iterative algorithm to solve the regularized minimization.
From this static solution, we finally construct in Section \ref{seq:ot-fm} the dynamical controller realizing the minimally dissipative gate, by combining generative modeling techniques from the flow-matching \cite{bengio_flow_2021, liu_flow_2022}, and score-matching \cite{hyvarinen_estimation_2005} literature.

\section{Optimal transport formulation of two-bit logical operations}
\label{seq:theory}

\subsection{Single bit-erasure} 

We begin by recalling the paradigmatic minimal bit erasure model proposed by Landauer~\cite{landauer_irreversibility_1961} in the formalism of stochastic thermodynamics~\cite{sekimoto_langevin_1998}.
In this model, a bit of information is represented by the sign of a single particle $x\in \RR$ relative to a fixed frame of reference.
Because the experimental realizations of bit erasure have focused on colloidal particles in the condensed phase environments, we describe the dynamics of the particle with an overdamped Langevin equation, subject to thermal fluctuations at temperature $T$ and to a time dependent driving force $f(x,t)$,
\begin{equation}
\label{eq:odlan}
    \der x_t = f(x_t,t)\der t + \sqrt{2 T}\der W_t,~~t\in [0,\tau],
\end{equation}
where $W_t$ is a one-dimensional Wiener process and the instantaneous distribution of $x$ is denoted $\rho_t(x)$.
Defining two disjoint sets $\CC_0 = \RR^-_*$ and $\CC_1 = \RR^+_*$ the particle's bit value is defined by its position in space. 
At time $t=0$, the particle position is Boltzmann distributed according to $\rho_0(x)$ associated to a symmetric double-well potential $U_0(x)=U_0(-x)$.
As such, its bit value is either 0 or 1 with equal probability.
The particle is driven from $t=0$ to $t=\tau$, at which point the potential is switched back to $U_0$ and the bit value is said to be completely erased if $\int_{\CC_0} \rho_\tau(x)\der x = 0$, i.e. there is no mass left in $\CC_0$ and the initial value of the bit cannot be recovered.

Following conventional thermodynamics \cite{sekimoto_stochastic_2010}, the work associated with the protocol decomposes into free energy and total entropy production terms:
\begin{equation}
\label{eq:work}
\begin{split}
    W &= \Delta F + \Delta S_{\rm tot}\\
    &= \int U_0(x)\left[\rho_\tau(x)-\rho_0(x)\right] +T \left[\int \rho_0(x) \log \rho_0(x) - \int \rho_\tau(x) \log \rho_\tau(x)\right] + \Delta S_{\rm tot}\\
    &=T \Dkl (\rho_\tau || \rho_0)+ \Delta S_{\rm tot}
\end{split}
\end{equation}
where $\Dkl(\nu || \mu) = \int \nu(\xb)\log(\nu(\xb)/\mu(\xb)) \der \xb$ denotes the Kullback-Leibler Divergence. 
Drawing on connections between stochastic thermodynamics and optimal transport \cite{aurell_refined_2012}, the thermodynamic speed-limit \cite{Saito2023} poses a tight lower bound on the work required to execute the transformation from $\rho_0$ to $\rho_\tau$ over the finite duration $\tau$:
\begin{equation}
    \label{eq:tsl}
    W \geq T \Dkl(\rho_\tau|| \rho_0) + \frac{\was^2(\rho_\tau, \rho_0)}{\tau}
\end{equation}
where the squared optimal transport, or 2-Wasserstein distance \cite{villani_optimal_2009} $\was^2(\nu,\mu)$ between distributions $\nu$ and $\mu$ is defined as the infimum 
\begin{equation}
    \label{eq:was}
    \was^2(\nu,\mu) = \underset{\pi \in \Pi(\nu,\mu)}{\inf} \int ||\xb - \yb||^2\pi(\xb, \yb)^2\der \xb \der \yb
\end{equation}
over all joint couplings $\pi \in \Pi(\nu,\mu)$ with prescribed marginals $\int \pi(\xb, \yb)\der \yb = \nu(\xb)$ and $\int\pi(\yb, \xb)\der \yb = \mu(\xb)$. In the bit erasure setup, some flexibility is allowed in $\rho_\tau$; denoting $\Rho$ the set of admissible distributions such that $\mu \in \Rho \iff \int_{\CC_0}\mu(\xb)\der \xb \equiv \mu(\CC_{0}) =0 $, the minimum work associated to the finite time erasure of one bit of information is simply
\begin{equation}
\label{eq:genbe}
    W^{\rm BE} = \underset{\rho_\tau \in \Rho}{\inf}\left[T \Dkl(\rho_\tau || \rho_0) + \frac{\was^2(\rho_\tau, \rho_0)}{\tau}\right].
\end{equation}
Importantly, we point out that the finite time Landauer principle as spelled out in \cite{Bechhoefer2020} is completely equivalent to the optimal transport formulation \eqref{eq:genbe} but specified to the one dimensional case. In this restricted setting, the Wasserstein distance admits a closed solution via a quadrature involving the cumulative distribution functions of the source and target distributions, and finding the optimal $\rho_\tau$ can be reduced to numerically solving a one dimensional ODE \cite{peyre_computational_2019}.
In the following, we show that the geometric formulation \eqref{eq:genbe} encompasses the more general case of multiple bit operations.

\subsection{Two bit logical operations} 

\begin{figure}
    \centering
    \includegraphics[width=0.8\linewidth]{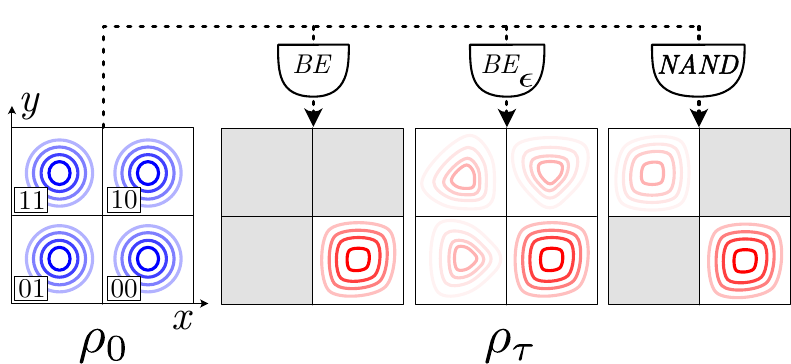}
    \caption{\footnotesize Schematics of various gate outputs. Initially, the mass is evenly spreat between the four quadrants, each corresponding to two bits of information. Measuring the particle's position at the end of the protocol yields the logical operation result. The grayed out zones carry no probability mass.}
    \label{fig:ops-schem}
\end{figure}

Following the previous discussion, we represent two bits of information as a single stochastic particle $\Xb \in \RR^2$ in contact with a thermal bath at temperature $T$. The initial state is symmetric both with respect to the $x$ and $y$ directions, and we define four bit sets as $\CC_{00} = \RR^+_* \times\RR^-_*$, $\CC_{01} = \RR^-_* \times\RR^-_*$,$\CC_{10} = \RR^+_* \times\RR^+_*$ and $\CC_{11} = \RR^-_* \times\RR^+_*$, depicted on the left-hand-side of Fig. \ref{fig:ops-schem}. 
A short calculation shows that the energetic computations \eqref{eq:work} do not depend on the dimensionality of the problem. 
In turn, the work associated to a logical gate is uniquely defined by the set $\Rho$ of constraints imposed on $\rho_\tau$. 

We thus examine the work associated with carrying out three distinct logical operations: complete two bit erasure, partial two bit erasure, and the NAND gate:
\begin{equation}
\label{eq:lg}
    \begin{split}
        &W^{\rm 2BE} = \underset{\rho_\tau \in \Rho^{\rm 2BE}}{\inf}\left[T \Dkl(\rho_\tau || \rho_0) + \frac{\was^2(\rho_\tau, \rho_0)}{\tau}\right];~\Rho^{\rm 2BE} = \left\{\rho; \rho(\CC_{00}) = 1 \right\}\\
        &W^{\rm 2BE}_\epsilon = \underset{\rho_\tau \in \Rho^{\rm 2BE}_\epsilon}{\inf}\left[T \Dkl(\rho_\tau || \rho_0) + \frac{\was^2(\rho_\tau, \rho_0)}{\tau}\right];~\Rho^{\rm 2BE}_\epsilon = \left\{\rho;\rho(\CC_{00}) = 1-\epsilon \right\}\\
        &W^{\rm NAND} = \underset{\rho_\tau \in \Rho^{\rm NAND}}{\inf}\left[T \Dkl(\rho_\tau || \rho_0) + \frac{\was^2(\rho_\tau, \rho_0)}{\tau}\right];~\Rho^{\rm NAND} = \left\{\rho; \rho(\CC_{11}) = 0.25,\rho(\CC_{00}) =  0.75  \right\},
    \end{split}
\end{equation}
and schematically depict candidate target distributions on the right-hand-side of Fig. \ref{fig:ops-schem}.
While the interpretation of $P^{\rm 2BE}$ and $P^{\rm 2BE}_\epsilon$ is straightforward, the NAND case merits comment. 
First, the target requirements ensure that at time $t=\tau$, the mass is supported on two diagonally opposed quadrants, indicating that the operation outcome can represent either a 0 or 1. 
Second, as in the one bit erasure protocol, observing the particle in $\CC_{00}$ at time $\tau$ does not allow for the reconstruction of the bit values at time $0$---the initial two-bit information has been partially erased, necessarily incurring energy cost.
Perhaps most importantly, note that OR and AND gates are identical to the NAND gate up to a redefinition of the final mass distribution.
Finally, similar to partial erasure set-ups \cite{Crutchfield2020, Wolpert2019, Bechhoefer2020_2}, there is a finite probability of logical error due to the stochastic nature of the particle dynamics, e.g. $\PP(\Xb_\tau \in \CC_{00} |\Xb_0 \in \CC_{00} ) > 0$. 
In order to mitigate these errors, standard signal processing techniques may need to be employed. 

Importantly, the logical gates \eqref{eq:lg} are purely formulated in terms of geometric distances and constraints on source and target distributions. 
As a result of this construction, it is both straightforward to extend them to higher number of bits (one particle in $n$ dimensions) or different numeral systems (base $n$ computations exploiting rotational symmetries of $\RR^2$).
While this is not the aim of this paper, the theoretical results and computational methods that we develop in the following sections---which naturally follow from \eqref{eq:lg}---could be expected to adapt to these cases.

\paragraph{Landauer bound} The quasi-static driving limit provides the most direct lower bound on gate dissipation. In analogy to the Landauer limit, for any gate $G$, associated constrained target set $\Rho^G$ and any $\tau$, the work is lower bounded by the purely information-geometric discrepancy
\begin{equation}
    W^{G} \geq \underset{\mu \in \Rho^{G}}{\inf}T \Dkl(\mu || \rho_0)\equiv {\rm LB^G}.
\end{equation}
yielding explicit Landauer bounds for quadrant symmetric source distributions reported in Table \ref{tab:landauer_bounds} of the Appendix \ref{app:landauer_bounds}.

\paragraph{Speed and accuracy limits} Beyond the quasi-static limit, the optimal transport formulation \eqref{eq:lg} provides a way to formalize intuitive energy considerations : faster and more accurate operations dissipate more.
The former statement is immediate :
both KL divergence and Wasserstein distance are positive, such that for any $\mu \in \Rho^G$

\begin{equation}
    \tau_1 \geq \tau_2 \implies T \Dkl(\mu||\rho_0) + \frac{\was^2(\mu, \rho_0)}{\tau_1} \leq T \Dkl(\mu||\rho_0) + \frac{\was^2(\mu, \rho_0)}{\tau_2}.
\end{equation}
In turn, it is immediate that
\begin{equation}
    \label{eq:speedlimit}
    \tau_1 \geq \tau_2 \implies W^{G}(\tau_1) \leq W^{G}(\tau_2).
\end{equation}
The accuracy limit is more involved, and we restrict the discussion to partial two bit erasure, namely we show that for every $\tau$

\begin{equation}
    \label{eq:accuracylimit}
    \epsilon_1 \geq \epsilon _2 \implies W^{2BE}_{\epsilon_1} \leq W^{2BE}_{\epsilon_2}.
\end{equation}
We first introduce the relaxed partial erasure gate
\begin{equation}
    \label{eq:relaxpe}
    W^{\rm 2BE}_{\leq \epsilon} = \underset{\rho_\tau \in \Rho^{\rm 2BE}_{\leq\epsilon}}{\inf}\left[T \Dkl(\rho_\tau || \rho_0) + \frac{\was^2(\rho_\tau, \rho_0)}{\tau}\right];~\Rho^{\rm 2BE}_{\leq\epsilon} = \left\{\rho;\rho(\CC_{00}) \geq 1-\epsilon \right\}
\end{equation}
associated to driving at least $1-\epsilon$ fraction of the total mass to $\CC_{00}$. 
For $\epsilon_2 \leq \epsilon_1$, $\Rho^{\rm 2BE}_{\leq\epsilon_2} \subseteq \Rho^{\rm 2BE}_{\leq\epsilon_1}$, such that $W^{\rm 2BE}_{\leq \epsilon_1} \leq W^{\rm 2BE}_{\leq \epsilon_2}$.  
In Appendix \ref{app:accuracylimit} we show that $W^{\rm 2BE}_{\leq \epsilon} = W^{\rm 2BE}_{\epsilon}$, yielding the accuracy limit

\begin{equation}
    \label{eq:accuracylimit}
    \epsilon_1 \geq \epsilon_2 \implies W^{\rm 2BE}_{\epsilon_1} \leq W^{\rm 2BE}_{\epsilon_2}.
\end{equation}
Finally, following the same arguments, it is easily seen that that for any gate $G$ with target support $\CC_{G}$ such that $\rho_\tau(\CC_{G}) = 1$, the minimal work associated to the partial gate $G_{\epsilon}$ with $\rho_{\tau}(\CC_{G}) = 1-\epsilon$ obeys the same accuracy limit
\begin{equation}
    \label{eq:accuracylimitgeneral}
    \epsilon_1 \geq \epsilon_2 \implies W^{G}_{\epsilon_1} \leq W^{G}_{\epsilon_2}.
\end{equation}
Together, these speed and accuracy tradeoffs Eq. \eqref{eq:speedlimit}, \eqref{eq:accuracylimitgeneral} outline optimization opportunities for realizing logical gates within fixed energy budgets. 
source distribution.

\paragraph{Logical gates in product systems} 
A natural question arises: does conducting two-bit erasure in two dimensions offer energetic advantages compared to manipulating two separate one-dimensional bits? 
While this remains unclear in general,  we demonstrate that for product source distributions $\rho_0(\Xb) = \rho^x_0(x)\rho^y_0(y)$, both approaches are \textit{energetically equivalent}. Abusing notations, we denote the instantaneous marginals
$\rho^x_t(x) =\int\rho_t(x,y)\der y$ and compute the two bit erasure work: 

\begin{equation}
\label{eq:impth1}
    \begin{split}
        W^{\rm 2BE} &= \underset{\rho_\tau \in \Rho^{\rm 2BE}}{\inf}\left[T \Dkl(\rho_\tau || \rho_0) + \frac{\was^2(\rho_\tau, \rho_0)}{\tau}\right]\\
        &= \underset{\rho_\tau \in \Rho^{\rm 2BE}}{\inf}\left[T \Dkl(\rho_\tau || \rho^x_\tau \rho^y_\tau) + T\int_{\RR^2} \rho_\tau(x,y)\log\left(\frac{\rho^x_\tau(x)\rho^y_\tau(y)}{\rho^x_0(x)\rho^y_0(y)}\right)\der x\der y +\frac{\was^2(\rho_\tau, \rho_0)}{\tau}\right]\\
        &= \underset{\rho_\tau \in \Rho^{\rm 2BE}}{\inf}\left[T \Dkl(\rho_\tau || \rho^x_\tau \rho^y_\tau) + T \Dkl(\rho^x_\tau || \rho^x_0) + T \Dkl(\rho^y_\tau || \rho^y_0 ) +\frac{\was^2(\rho_\tau, \rho_0)}{\tau}\right].
    \end{split}
\end{equation}
The mutual information $\Dkl(\rho_\tau || \rho^x_\tau \rho^y_\tau)$ of $\rho_{\tau}$ is always positive, such that 

\begin{equation}
\label{eq:impth2}
        W^{\rm 2BE} \geq \underset{\rho_\tau \in \Rho^{\rm 2BE}}{\inf}\left[T \Dkl(\rho^x_\tau || \rho^x_0) + T \Dkl(\rho^y_\tau || \rho^y_0 ) +\frac{\was^2(\rho_\tau, \rho_0)}{\tau}\right]\\.
\end{equation}
Using the definition \eqref{eq:was} of the Wasserstein distance, and denoting $\pi_\tau(x_\tau, y_\tau, x_0, y_0)$ the optimal coupling associated to $\rho_\tau$ and $\rho_0$, we obtain

\begin{equation}
\label{eq:impth2}
\begin{split}
        W^{\rm 2BE} &\geq \underset{\rho_\tau \in \Rho^{\rm 2BE}}{\inf}\left[T \Dkl(\rho^x_\tau || \rho^x_0)  + \tau^{-1}\int_{\RR^2}|x_\tau - x_0|^2\pi_\tau(x_\tau, x_0)\der x_\tau \der x_0\right.\\
        &\left.\hspace{50pt}+ T \Dkl(\rho^y_\tau || \rho^y_0 ) + \tau^{-1}\int_{\RR^2}|y_\tau - y_0|^2\pi_\tau(y_\tau, y_0)\der y_\tau \der y_0\right]\\
        &\geq \underset{\rho_\tau \in \Rho^{\rm 2BE}}{\inf}\left[T \Dkl(\rho^x_\tau || \rho^x_0)  + \tau^{-1}\int_{\RR^2}|x_\tau - x_0|^2\pi_\tau(x_\tau, x_0)\der x_\tau \der x_0 \right]  \\
        &~~~+\underset{\rho_\tau \in \Rho^{\rm 2BE}}{\inf}\left[T \Dkl(\rho^y_\tau || \rho^y_0 ) + \tau^{-1}\int_{\RR^2}|y_\tau - y_0|^2\pi_\tau(y_\tau, y_0)\der y_\tau \der y_0\right]\\
        &\geq \underset{\rho^x_\tau \in \Rho^{\rm 1BE}}{\inf}\left[T \Dkl(\rho^x_\tau || \rho^x_0)  + \tau^{-1}\was^2(\rho_\tau^x, \rho_0^x)\right]\\
        &~~~+\underset{\rho^y_\tau \in \Rho^{\rm 1BE}}{\inf}\left[T \Dkl(\rho^y_\tau || \rho^y_0)  + \tau^{-1}\was^2(\rho_\tau^y, \rho_0^y)\right]
    \end{split}
\end{equation}
where the second inequality derives from the $\inf$ operation, and the third inequality holds because $\rho_\tau \in \Rho^{\rm 2BE} \implies \rho_\tau^x \in \Rho^{\rm1BE}$. As a result, we obtain the following lower bound for factorized source distribution :

\begin{equation}
    \label{eq:impth3}
    W^{\rm 2BE}(\rho_0^x\rho_0^y) \geq W^{\rm 1BE}(\rho_0^x) + W^{\rm 1BE}(\rho_0^y).
\end{equation}
It now suffices to show that the inequality can be saturated, which is merely done by constructing the driving force from the one dimensional optimal bit erasure controllers :$f^{\rm 2BE}(x,y) = (f^{\rm 1BE}(x), f^{\rm 1BE}(y))$. The resulting target distribution is the product of the two optimal one bit target distributions $\rho_\tau(x,y) = \rho^x_\tau(x)\rho^y_\tau(y)$, and

\begin{equation}
    \label{eq:impth3}
    W^{\rm 2BE}(\rho_0^x\rho_0^y) = W^{\rm 1BE}(\rho_0^x) + W^{\rm 1BE}(\rho_0^y).
\end{equation}
The factorized erasure theorem \eqref{eq:impth3} thus establishes that the finite-time Landauer limits remain inevitable even when operating in higher dimensions. While it is tempting to extrapolate to partial two-bit erasure, we prove in Appendix \ref{app:partialerasureimpth} that it is actually not the case. Importantly, we emphasize that this results relies on two assumptions : the source distribution is factorized, namely the system is prepared by thermalization in a potential $U_0(\Xb) = U_0^x(x) + U_0^y(y)$, and the optimal target distribution can be factorized.
However, both assumptions are restrictive.
First, multiple mode source distributions can take any non-product mixture form $\rho_0(\Xb) = \sum_i \rho_0(\Xb - \Xb_i)$. 
Second, logical gates, and in particular the NAND gate, do not necessarily allow for factorized optimal target distributions.
We thus propose an algorithmic route to constructing minimally dissipative logical gates for arbitrary source distributions.  

\section{Regularized logical gates}
\label{seq:regularized-lg}

In the following we consider $G$ to be one of the three gates in \eqref{eq:lg}, $\rho_0$ a fixed initial state, and recall the associated work, explicitly writing the Wasserstein distance as infimum over couplings:
\begin{equation}
    \label{eq:genericgate}
    W^{G} = \underset{\rho_\tau \in \Rho^G}{\inf}\left[T \Dkl(\rho_\tau||\rho_0) + \tau^{-1}\underset{\pi \in \Pi(\rho_\tau, \rho_0)}{\inf}\int ||\Xb_\tau - \Xb_0||^2\pi(\Xb_\tau, \Xb_0)\der\Xb_\tau\der\Xb_0\right].
\end{equation}
We now proceed to exchange the two $\inf$ operations. Denoting as a shorthand the marginals $\pi(\Xb_\tau \times \RR^2) = \pi_\tau(\Xb_\tau)$,$\pi(\RR^2 \times \Xb_0) = \pi_0(\Xb_0)$ we obtain the unconstrained coupling minimization problem
\begin{equation}
\label{eq:couplinggate}
    W^{G} = \underset{\pi}{\inf}\left[\iota_{\rho_0}(\pi_0) + T \Dkl(\pi_\tau || \rho_0) + \iota_{G}(\pi_\tau) + \tau^{-1}\int ||\Xb_\tau - \Xb_0||^2\pi(\Xb_\tau, \Xb_0)\der\Xb_\tau\der\Xb_0\right],
\end{equation}
where the indicator functions
\begin{equation}
\label{eq:iota}
\iota_{\rho_0}(\mu)=\left\{\begin{split}
&0\ \text{if}\ \mu = \rho_0 \\
&+\infty\ \text{else}
\end{split}\right.
\hspace{10pt}
\iota_{G}(\mu)=\left\{\begin{split}
&0\ \text{if}\ \mu \in \Rho^G \\
&+\infty\ \text{else}
\end{split}\right.
\end{equation}
impose the marginal constraints associated to the gate $G$. 
The coupling minimization problem \eqref{eq:couplinggate}, reminiscent of the Monge-Kantorovitch definition of optimal transport \cite{benamou_mongekantorovitch_2002}, provides geometrical insight into logical gate energetics but presents computational challenges. 
Even in standard optimal transport problems with prescribed marginals, solutions are generally neither analytically nor computationally tractable. 
However, recent advances in computational optimal transport \cite{cuturi_sinkhorn_2013,Peyre2017,cuturi_semidual_2018} employ regularization methods to efficiently obtain optimal couplings with controllable approximation, paving the way for near-optimal gate design.

\subsection{Entropic regularization}

Inspired by the celebrated Sinkhorn algorithm \cite{sinkhorn1964relationship} for optimal transport, we consider the optimal regularized work 

\begin{equation}
\label{eq:regwork}
    W^{G}_\epsilon = T \Dkl(\pi^\epsilon_\tau || \rho_0) + \tau ^{-1} \int ||\Xb_\tau - \Xb_0||^2\pi^\epsilon(\Xb_\tau, \Xb_0)\der\Xb_\tau\der\Xb_0
\end{equation}
where $\pi^\epsilon$ is the arginf of the regularized objective
\begin{equation}
    \label{eq:regobj1}
    \pi^\epsilon = {\rm arginf}\left[\iota_{\rho_0}(\pi_0) + T \Dkl(\pi_\tau || \rho_0) + \iota_{G}(\pi_\tau) + \tau^{-1}\int ||\Xb_\tau - \Xb_0||^2\pi(\Xb_\tau, \Xb_0)\der\Xb_\tau\der\Xb_0 + \epsilon H(\pi)\right],
\end{equation}
with $H(\pi) = \int \pi( \Xb_\tau,  \Xb_0))\log(\pi(\Xb_\tau, \Xb_0))\der \Xb_\tau \der \Xb_0$ the entropy of the coupling. Following conventions \cite{Chizat2018} and to simplify notations, we introduce the transport cost function $c(\Xb_0, \Xb_\tau) = \tau^{-1}||\Xb_0 - \Xb_\tau||^2$ and scalar product $\langle \pi , c\rangle = \int \pi.c$ such that the regularized work and coupling can be rewritten as

\begin{equation}
    \label{eq:regobj2}
    \begin{split}
        &\pi^\epsilon = {\rm arginf}\left[\iota_{\rho_0}(\pi_0) + T \Dkl(\pi_\tau || \rho_0) + \iota_{G}(\pi_\tau) + \epsilon \Dkl(\pi||e^{-c/\epsilon})\right],\\
        &W^{G}_\epsilon = T \Dkl(\pi_\tau^\epsilon ||\rho_0) + \langle \pi^\epsilon, c\rangle. 
    \end{split}
\end{equation}
The regularized formulation \eqref{eq:regobj2} is the main contribution of this paper as it recasts the stochastic thermodynamics of logical gates as a non-standard specific instance of regularized unbalanced optimal transport (UOT)\cite{liero_optimal_2018}. 
As a result, we adapt UOT arguments to derive important physical and practical implications.

First, using the respective minimizing properties of the regularized $\pi^\epsilon$
and optimal $\pi^G$ couplings yields
\begin{equation}
    \label{eq:regobjconv}
    \begin{split}
        &T \Dkl(\pi^\epsilon_\tau||\rho_0) + \langle \pi^\epsilon, c\rangle + \epsilon H(\pi^\epsilon) \leq  T \Dkl(\pi^{G}_\tau||\rho_0) + \langle \pi^{G}, c\rangle + \epsilon H(\pi^{G}),\\
        &T \Dkl(\pi^\epsilon_\tau||\rho_0) + \langle \pi^\epsilon, c\rangle  \geq  T \Dkl(\pi^{G}_\tau||\rho_0) + \langle \pi^{G}, c\rangle,\\
        & \implies 0 \leq W^G_\epsilon - W^G \leq \epsilon(H(\pi^{G}) - H(\pi^\epsilon) ).
    \end{split}
\end{equation}
While the regularized work is provably suboptimal, the excess energy linearly converges to 0 in the small $\epsilon$ limit (we here assume finiteness of entropies, which from a applied perspective is always true).
Furthermore, this energetic waste is a small price to pay as entropic regularization offers exponential computational speed-ups for protocol determination.
Indeed, rewriting the regularized coupling \eqref{eq:regobj2} in the more compact form
\begin{equation}
    \pi^\epsilon = {\rm arginf}\left[F_0(\pi_0) + F_\tau^{G}(\pi_\tau) + \epsilon \Dkl(\pi||e^{-c/\epsilon})\right],\ F_0(\mu) = \iota_{\rho_0}(\mu),\ F_\tau(\mu) = T \Dkl(\mu || \rho_0) + \iota_{G}(\mu)
\end{equation}
reveals that both 
constraint functions $F_0$ and $F_\tau$ are convex lower semicontinuous and thus fall within the applicability scope of recently developed fast UOT iterative algorithms to construct $\pi^\epsilon$ \cite{Chizat2018}.

\subsection{Iterative algorithms} In the following, we closely follow results and notation from \cite{Chizat2018} and only reproduce formulae useful in our context. Denoting $K = e^{-c / \epsilon}$, the regularized coupling is known \cite{rockafellar1967duality} to take the form
\begin{equation}
    \label{eq:uot1}
    \pi^\epsilon(\Xb_\tau, \Xb_0) = e^{\frac{u(\Xb_\tau)}{\epsilon}}K(\Xb_\tau, \Xb_0)e^{\frac{v(\Xb_0)}{\epsilon}},
\end{equation}
with unknown dual functions $u,v : \RR^2 \to \RR$. Denoting further the kernel convolution $(K*s)(\Xb_0) = \int K(\Xb_0, \Xb_\tau)s(\Xb_\tau)\der \Xb_\tau$ The dual iterative scheme 

\begin{equation}
    \label{eq:itscheme}
    \begin{split}
    &u^{\ell +1} = \underset{u}{\rm argmax}\left[-\Tilde{F}_0(-u) - \epsilon \langle e^{\frac{u^\ell}\epsilon},(K*e^{\frac{v^\ell}{\epsilon}})\rangle\right]\\
    &v^{\ell +1} = \underset{v}{\rm argmax}\left[-\Tilde{F}^G_\tau(-v) - \epsilon \langle e^{\frac{v^\ell}\epsilon},(K^T*e^{\frac{u^{\ell+1}}{\epsilon}})\rangle\right]\\
    \end{split}
\end{equation}
where the Legendre Transform $\Tilde{F}$ is defined as 

\begin{equation}
    \label{eq:legendre}
    \Tilde{F} = \underset{p}{\rm sup}\left[\langle p,u\rangle - F(p)\right],
\end{equation}
converges towards $\pi^\epsilon$ \cite{linial1998deterministic}. In turn, provided the updates are analytically tractable, the iterative scheme constructs the regularized coupling associated to gate $G$. For clarity, we include these computations in the case of complete two bit erasure in the main text and relegate partial erasure and NAND gate to the Appendix \ref{app:dualupdate}.

Focusing first on the $u$ update, we compute the Legendre transform
\begin{equation}
\label{eq:cbeuLT}
    \Tilde{F}_0(u) = \underset{p}{\rm sup}\left[\langle p,u\rangle - \iota_{\rho_0}(p)\right] = \langle \rho_0,u\rangle.
\end{equation}
Plugging into the update \eqref{eq:itscheme} and zeroing out the first variation yields the fixed marginal update
\begin{equation}
\label{eq:cbeuupdate}
\frac{\partial}{\partial u}\left[\langle \rho_0,u\rangle - \epsilon \langle e^{-\frac{u}{\epsilon}},(K*e^{\frac{v^{\ell}}{\epsilon}})\rangle \right] = 0 \iff \rho_0 = e^{\frac{u}{\epsilon}}(K*e^{\frac{v^{\ell}}{\epsilon}}) \iff u^{\ell+1} = \epsilon\left(\log(\rho_0) - \log((K*e^{\frac{v^{\ell}}{\epsilon}}))\right).
\end{equation}
We next consider the $v$ update associated to the Legendre transform

\begin{equation}
\label{eq:cbevLT}
    \Tilde{F}_\tau(v) = \underset{p}{\rm sup}\left[\langle p ,v\rangle - T \Dkl(p||\rho_0) + \iota_{\rm 2BE}(p)\right]
\end{equation}
where the indicator function enforces $p(\CC_{00}) = 1$. Taking again a first variation yields 
\begin{equation}
\label{eq:cbevLT2}
    p = \rho_0 e^{\frac{v}{T}}\mathbbm{1}_{\overline{\CC_{00}}} \implies \Tilde{F}_\tau(v) = T \langle \rho_0,  (e^{\frac{v}{T}}\mathbbm{1}_{\overline{\CC_{00}}} - 1) \rangle.
\end{equation}
Finally, combining the Legendre transform \eqref{eq:cbevLT2} together with the update scheme \eqref{eq:itscheme}, we obtain

\begin{equation}
\label{eq:cbevupdate}
    v^{\ell +1} = \frac{T \epsilon}{T + \epsilon}\left[\log(\rho_0 \mathbbm{1}_{\overline{\CC_{00}}}) - \log(K*e^{\frac{u^{\ell + 1}}{\epsilon}})\right]
\end{equation}
where, by convention, $\log(0) = -\infty$, such that $\Xb \in \overline{\CC_{00}} \implies \pi_\tau(\Xb) =0$. 
The dual updates \eqref{eq:cbeuupdate} and \eqref{eq:cbevupdate} hence provide a straightforward way to iteratively construct $\pi^\epsilon$ given only the data $\rho_0$ and cost function $c$. 
Note that while the updates associated to the partial erasure and NAND gate provided in Appendix \ref{app:dualupdate} take slightly different forms, the underlying logic is strictly equivalent.

\subsection{Numerical illustrations}

The regularized formulation \eqref{eq:couplinggate} provides both arbitrarily precise approximation of minimal work for computational logical gates and a straightforward algorithmic procedure to obtain it, extending well beyond existing one-dimensional methods.
Although the iterative algorithm's convergence rate is linear in $\epsilon$ \cite{knight2008}, our numerical experiments reveal that reasonably large epsilon values yield excellent approximations, offering an attractive balance between computational efficiency and precision.

\begin{figure}[htbp]
    \centering
    
    % First subfigure
    \begin{subfigure}[t]{0.45\textwidth}
        \centering
        \includegraphics[width=\textwidth]{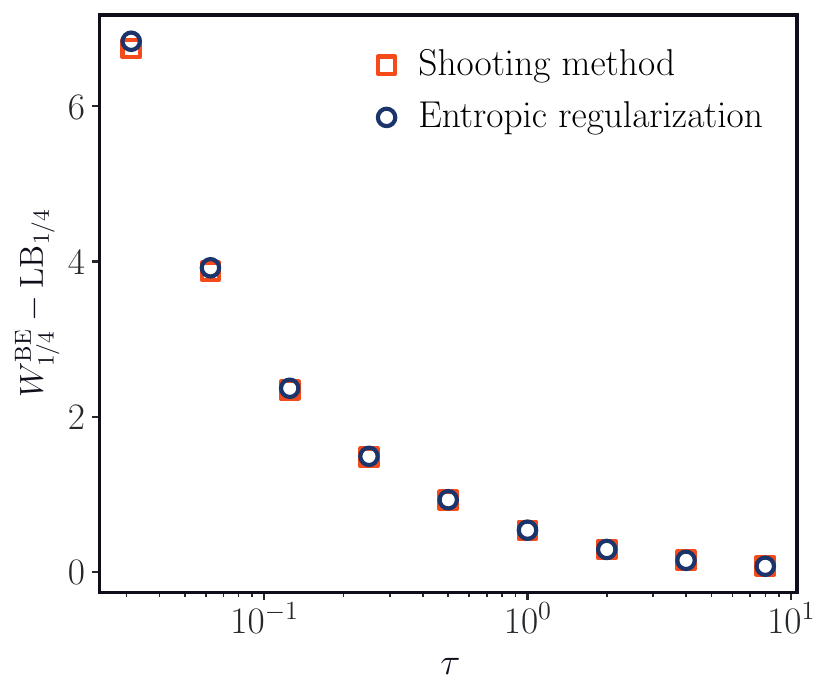}
        \caption{\footnotesize One dimensional partial bit erasure gate associated to a double well potential $U_0(x) = 4.0 ((x/1.043)^2-1)^2$. The circle data corresponds to the regularized work \eqref{eq:couplinggate} for $\epsilon=5.10^{-3}$, in agreement with the square data obtained using the shooting method from \cite{Bechhoefer2020_2}. The lower bound ${\rm LB}_{1/4}$ corresponds to the infinite time driving limit.
        }
        \label{fig:be1}
    \end{subfigure}
    \hfill
    % Second subfigure
    \begin{subfigure}[t]{0.45\textwidth}
        \centering
        \includegraphics[width=\textwidth]{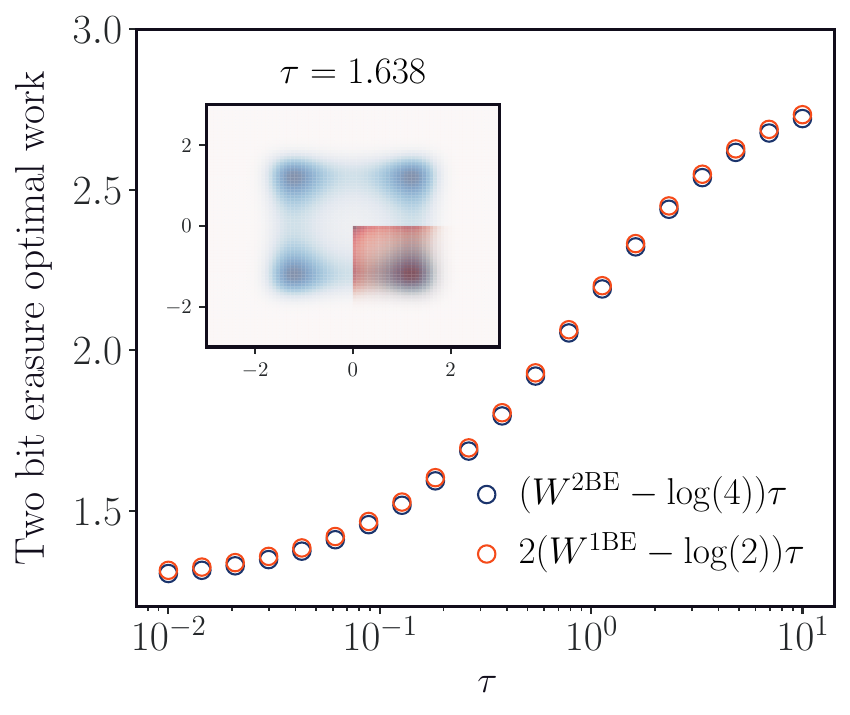}
        \caption{\footnotesize Optimal factorized two bit erasure work and associated one bit erasure work with regularization $\epsilon=10^{-2}$. The bit erasure theorem \eqref{eq:impth3} is numerically verified.\textbf{(Inset)} Superposed source (blue) and target (red) histograms for a given $\tau$ value.}
        \label{fig:be2}
    \end{subfigure}
    \centering
    %\caption{One and two dimensional bit erasure gates.}
    \label{fig:two_panels}
\end{figure}

\begin{figure}[!htbp]
    \centering
    \includegraphics[width=0.45\textwidth]{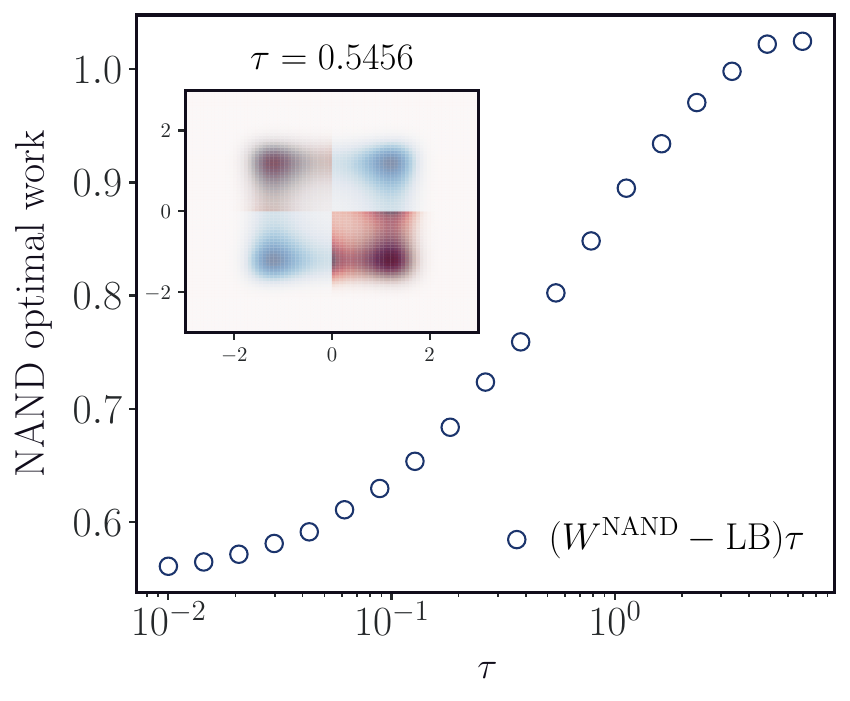}
    \caption{Regularized Minimal work associated to the NAND gate for the factorized potential $U_0(\Xb) = U_0^x(x)+U_0^y(y)$, 
    using $n_{\rm bins} = 80$ and  $\epsilon = 10^{-2}$. The Landaeur lower bound ${\rm LB} =\frac{3}{4}\log(3)$ corresponds to the infinite driving time limit.
    \textbf{(Inset)} Superposed source (red) and target (blue) marginals for a given $\tau$ value. }
    \label{fig:nand}
\end{figure}

We first validate our computational method by recovering one-dimensional partial bit erasure results \cite{Bechhoefer2020_2} and comparing with our iterative method in Fig. \ref{fig:be1}. We then consider the two-dimensional case of complete bit erasure associated to a factorized quadruple well potential $U_0(\Xb)=U_0^x(x) + U_0^y(y)$. The associated logical gate satisfies the assumptions of the factorized bit erasure theorem \eqref{eq:impth3} thus providing a self-consistent check of the method, shown on Fig. \ref{fig:be2}.
Note that in both cases, the regularized work closely agrees with the \emph{unregularized} work, highlighting that the method remains robust even for moderate values of $\epsilon$.
\noindent We turn next to the factorized potential NAND gate, for which the target constraints ensure that no one-dimensional based solutions exists. 
Carrying out the corresponding iterative scheme \eqref{eq:itscheme}, we characterize on Fig. \ref{fig:nand} the $\tau$ dependence of the associated minimal work, recovering the $\tau$ monotonicity \eqref{eq:speedlimit}.
These numerical results highlight the relevance of optimal transport tools in determining minimal energy costs associated to information processing. However, realizing such minimally dissipative operations requires more than the static coupling obtained by solving the regularized minimization \eqref{eq:couplinggate}. In the following, we extend the dynamical framework first proposed in \cite{KR25_1} to the case of logical operations.

\section{Optimal controller design}
\label{seq:ot-fm}

\subsection{Building protocols with optimal transport flow matching}
We hereafter take $\tau = 1$ for simplicity.
While the definitions of the various logical gates \eqref{eq:lg} imply that the target distribution $\rho_1$ is unknown,
computing the regularized optimal coupling $\pi^{\epsilon}$ in Eq. \eqref{eq:regobj1} fixes the target distribution as one of its marginals $\rho_1 = \pi^\epsilon_1$.
As a result, the KL term $T \Dkl(\rho_1 || \rho_0)$ becomes a constant, and minimizing the work reduces to a minimization over all distribution trajectories $\{\rho_t\}_{0\leq t \leq 1}$ satisfying both fixed source and target constraints. 
Building on connections between optimal transport and stochastic thermodynamics first uncovered in \cite{aurell_refined_2012}, it is known that that the minimally dissipative trajectory is the constant speed Wasserstein geodesic \cite{santambrogio_optimal_nodate} defined by its marginals
\begin{equation}
\label{eq:geodesic}
    \Xb_t\sim \rho^*_t\ \text{where}\ \Xb_t =  (1-t) \Xb_0 + t \Xb_1\ \text{with}\ (\Xb_0,\Xb_1) \sim \pi^{\epsilon},
\end{equation}
with $\pi^{\epsilon}$ the regularized optimal coupling \eqref{eq:was}.
Consequently, designing an optimal gate amounts to constructing control forces that keep the Langevin system \eqref{eq:odlan} on the Wasserstein geodesic at all times, up to the $\epsilon$-entropic approximation error. 

In the absence of thermal fluctuations, the system subject to control forces $f$ evolves according to the flow equation $\partial_t \rho_t = -\nabla\left[f.\rho_t\right]$.
Such dynamical formulations strongly echo continuous normalizing flows generative modeling methods \cite{chen_neural_2018}, which generate samples from a target distribution $\rho_1$ by propagating source samples from a base distribution $\rho_0$ via a learnable ODE $\dot{\Xb}_t = f_{\theta}(\Xb_t,t)$. 
Furthermore, recent works in high dimensional  sampling tasks (e.g. images) \cite{liu_flow_2022} have identified heuristic progress by imposing ``straight path flows''.
Specifically, the optimal transport flow matching strategy \cite{tong_improving_2023} (OTFM) proposes to represent the flow field $f_\theta$ as the minimizer of the flow matching objective

\begin{equation}
    \label{eq:otfm1}
    f^* = \underset{f_\theta}{\rm arginf}\  \mathbb{E}\left[||f_\theta((1-t)\Xb_0 + t \Xb_1,t) - (\Xb_1-\Xb_0)||^2\right];\ (\Xb_0, \Xb_1)\sim \pi^{\epsilon}, t \sim \mathcal{U}[0,1].
\end{equation}
Importantly, the resulting dynamical system satisfies both source and target constrains while flowing along the Wasserstein geodesic.
In the case of a logical gate $G$, the identification of $\pi^{\epsilon}_G$ using the iterative procedure described above allows for easy approximate multinomial sampling of pairs $(\Xb_0,\Xb_1)$ and thus training of the corresponding noiseless flow field.

In the Stochastic Thermodynamics paradigm, the thermal fluctuations resulting from interactions with the bath degrees of freedom steer the system away from its optimal course by adding a diffusion term

\begin{equation}
    \label{eq:thermalpde}
    \partial_t \rho_t = -\nabla\left[f^* \rho_t\right] + T \Delta \rho_t,
\end{equation}
such that $\rho_t = \rho^*_t * \mathcal{N}(0, 2 T t)$. To mitigate these diffusion effects, we introduce an additional control term approximating the score of the optimal trajectory $s_\theta(\Xb,t) \simeq \nabla \log \rho^*_t(\Xb)$. Provided the controller $s_\theta$ is exact, it is easy to see that the dynamical system
\begin{equation}
    \label{eq:finalds}
    \der \Xb_t = \left[f^*(\Xb,t) + T s_\theta(\Xb_t,t)\right] \der t + \sqrt{2 T}\der \Wb_t
\end{equation}
follows a Wasserstein geodesic and thus realizes the optimal logical gate. In the small dimensional setting associated to the gates discussed in the previous sections, (2+1) grid-based approximations of $\nabla \log \rho^*_t$ are simply obtained by populating instantaneous histograms of $\rho_t^*$ from the static coupling $\pi^\epsilon$. Note, however, that in higher dimensional systems, a number of score-matching approximations, widely successful in generative diffusion models, could be employed \cite{hyvarinen_estimation_2005,ho_denoising_2020}. 

\subsection{Numerical illustration}

Approximate optimal gate design thus reduces to two distinct steps. First, the static coupling minimization problem \eqref{eq:couplinggate} is entropically regularized and iteratively solved to obtain the approximate coupling $\pi^\epsilon$. In turn, geodesic samples are generated at all times $t\in [0,1]$ to construct the optimal controller comprising of the noiseless optimal flow field $f^*$ \eqref{eq:otfm1} and the instantaneous score $s_\theta$. 

\begin{figure}[htbp]
    \centering
    \includegraphics[width=0.45\textwidth]{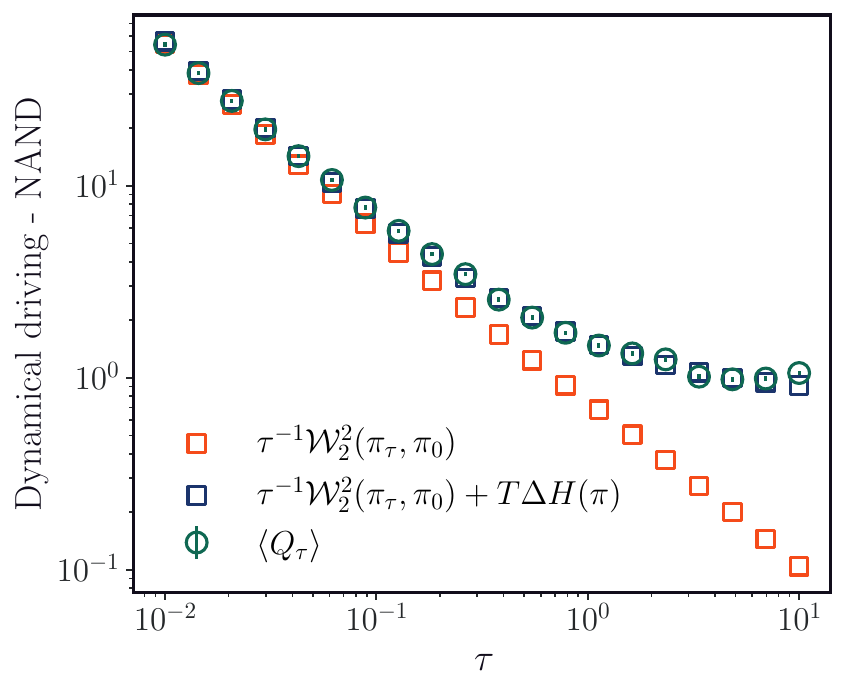}
    \caption{\footnotesize Empirical dissipated heat from Langevin simulations of \eqref{eq:finalds}. The optimal transport curve corresponds to $\was^2(\rho_\tau, \rho_0)\tau^{-1}$, whereas the optimal heat curve accounts for the additional entropic contribution \eqref{eq:dissipatedheat}.}
    \label{fig:screentaunand}
\end{figure}

We illustrate this workflow on the non-trivial NAND case.
As an empirical measure of dissipation, we monitor the time-dependent mean cumulative dissipated heat 

\begin{equation}
    Q_t = \mathbbm{E}\left[\int_0^t (f_\theta(\Xb_s,s) + T s_\theta(\Xb_s,s))\circ \der \Xb_s \right]
\end{equation} 
whose optimal value is given by

\begin{equation}
\label{eq:dissipatedheat}
    Q^*_\tau = \tau^{-1}\was^2(\rho_\tau, \rho_0 ) + T \left[H(\rho_\tau) - H(\rho_0)\right].
\end{equation}
As shown on Fig. \ref{fig:screentaunand}, the dynamical dissipation agrees reasonably well with the optimal static prediction in both the small $\tau$ transport-dominated and large $\tau$ KL-dominated regimes.

\begin{figure}[!htbp]
    \centering
    \includegraphics[width=0.8\textwidth]{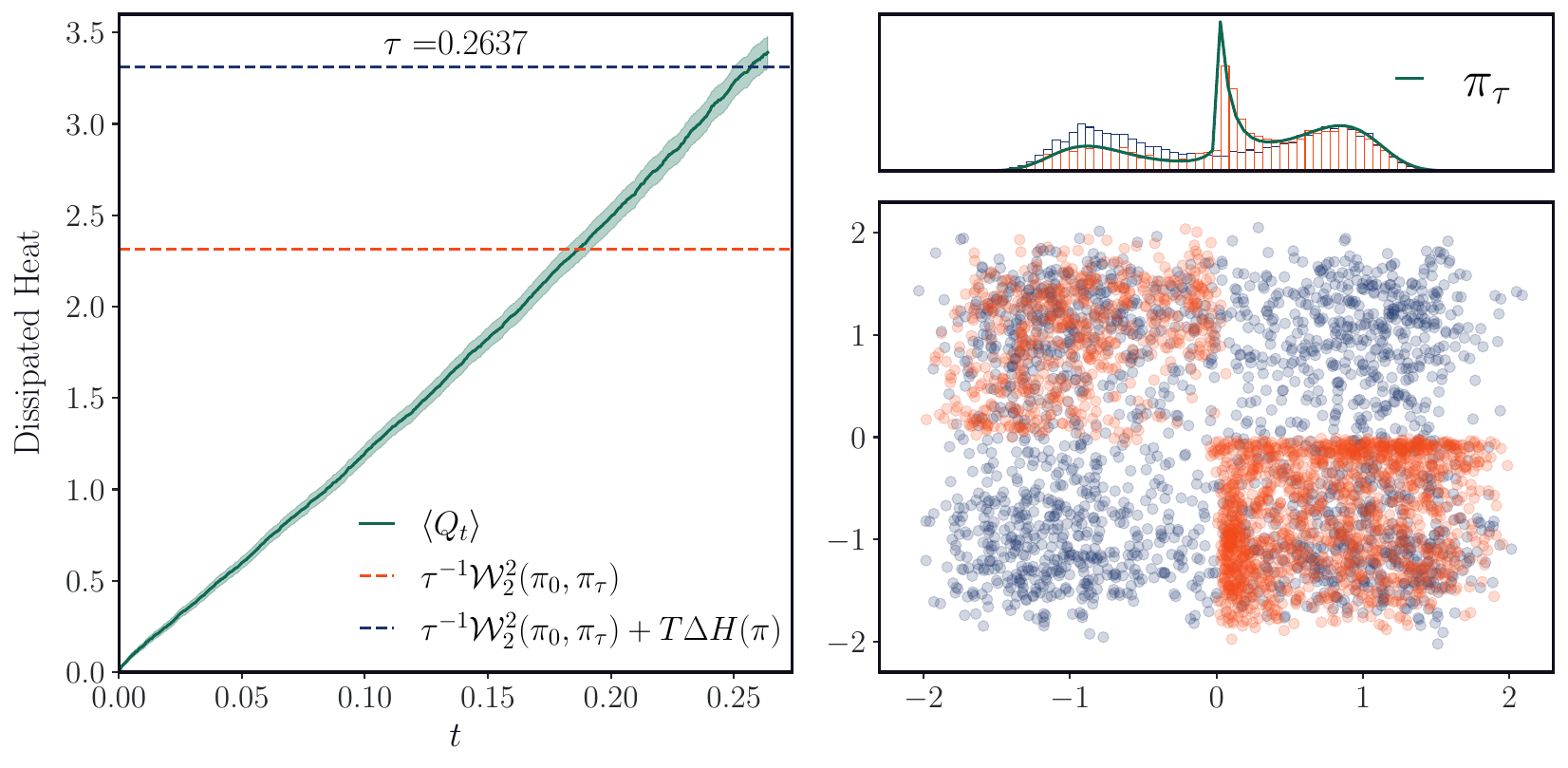} 
    \caption{\footnotesize Single $\tau=0.2637$ driving. \textbf{(Left)} Mean cumulative dissipated heat compared to the static predictions from Fig. \ref{fig:screentaunand}. \textbf{(Right)} Source and target cloud points. The marginal histograms are compared to the $\pi^\epsilon$ coupling marginals for $\epsilon = 10^{-2}$}
    \label{fig:nanddynamical}
\end{figure}

\noindent Moreover, the target distribution is clearly realized with satisfying accuracy, as displayed on Fig. \ref{fig:nanddynamical}, validating the dynamical methodology.

\newpage
\section{Conclusion}

While the deep connections between the mathematical theory of optimal transport and minimally dissipative control problems have been appreciated for some time, it has proven difficult to utilize this connection in non-trivial problems due to the computational complexity of high-dimensional optimal transport.
These limitations have created bottlenecks to progress on crucial problems, including the design of thermodynamically optimal information processors.
Here, we recast the problem of manipulating bits of information in such a way that execute logical operations with minimum heat dissipation as unbalanced optimal transport problems.
Our approach extends beyond the classical one-dimensional Landauer principle to address multi-bit logical gates—complete erasure, partial erasure, and NAND operations—in higher-dimensional settings.
Furthermore, it is computationally efficient, controllably precise, and we believe it to be scalable beyond what we have shown here. 

This point of view and the resulting of optimal protocols for bitwise logical operations yield fundamental physical insight. First, we formalize intuitive energy-speed-accuracy trade-offs for logical operations, demonstrating that faster and more accurate operations necessarily dissipate more energy. Second, we showed that for factorized source distributions, two-dimensional bit erasure offers no energetic advantage over separate one-dimensional operations, though, interestingly, this equivalence breaks down for more general initial states and gate constraints.

In addition, this framework is extensible to a large class of physical problems: we provide a practical computational solution, combining entropic regularization of optimal transport with modern generative modeling techniques. The iterative algorithm for solving regularized coupling problems, together with flow matching and score matching methods, provides a concrete path toward implementing optimal controllers. 
These controllers, realizable in colloidal control experimental devices, bridge the gap between theoretical bounds and experimental implementation.

Ultimately, this work provides both fundamental insights into the thermodynamic limits of computation and practical tools for approaching these limits in real systems. 
As quantum and classical computing architectures make a push for efficiency, principled approaches to the design of computational platforms will become increasingly crucial for next-generation technologies.

\newpage

\appendix

\section{Landauer Bounds}\label{app:landauer_bounds}

In this section we report the quasistatic information theoretic Landauer bounds for each gate of the main text. These explicit expressions hold in the case of quadrant symmetric source distributions and are obtained by zeroing out the first variation of the the KL divergence over the constrained target distribution set $P^G$ :

\begin{table}[h]
\centering
\begin{tabular}{|l|c|}
\hline
\textbf{Gate} & \textbf{Landauer Bound} \\
\hline
$\mathrm{LB^{2BE}}$ & $\log(4)$ \\
\hline
$\mathrm{LB_\epsilon^{2BE}}$ & $(1-\epsilon)\log(1-\epsilon) + \epsilon \log(\epsilon) + \log(4) - \epsilon \log(3)$ \\
\hline
$\mathrm{LB^{NAND}}$ & $\frac{3}{4}\log(3)$ \\
\hline
\end{tabular}
\captionsetup{justification=centering}
\caption{\footnotesize Quasistatic Landaeur bounds.}
\label{tab:landauer_bounds}
\end{table}

\section{Accuracy Limit} \label{app:accuracylimit}

In this section, we come back to the accuracy limit \eqref{eq:accuracylimitgeneral}, and show that $W^{\rm 2BE}_{\leq \epsilon} = W^{\rm 2BE}_{\epsilon}$ where

\begin{equation}
    \begin{split}
        &W^{\rm 2BE}_{\epsilon} = \underset{\rho_\tau \in \Rho^{\rm 2BE}_{\epsilon}}{\inf}\left[T \Dkl(\rho_\tau || \rho_0) + \frac{\was^2(\rho_\tau, \rho_0)}{\tau}\right]\equiv\underset{\rho_\tau \in \Rho^{\rm 2BE}_{\epsilon}}{\inf}f(\rho_\tau);~\Rho^{\rm 2BE}_{\epsilon} = \left\{\rho;\rho(\CC_{00}) = 1-\epsilon \right\}\\
        &W^{\rm 2BE}_{\leq \epsilon} = \underset{\rho_\tau \in \Rho^{\rm 2BE}_{\leq\epsilon}}{\inf}\left[T \Dkl(\rho_\tau || \rho_0) + \frac{\was^2(\rho_\tau, \rho_0)}{\tau}\right]\equiv\underset{\rho_\tau \in \Rho^{\rm 2BE}_{\leq \epsilon}}{\inf}f(\rho_\tau);~\Rho^{\rm 2BE}_{\leq\epsilon} = \left\{\rho;\rho(\CC_{00}) \geq 1-\epsilon \right\}.
    \end{split}
\end{equation}
It is clear that $P^{2BE}_{\leq \epsilon}$ is a closed convex set on the appropriate space of probability measures, that $P^{2BE}_{\epsilon} = \partial P^{2BE}_{\leq \epsilon}$ and that $W_{\leq \epsilon}^{2BE}$ is strictly decreasing as a function of $\epsilon$.
Let us further denote $\rho^*_\epsilon$, $\rho^*_{\leq \epsilon}$ the respective arginf of the above given minimization problems. \\

\noindent Assume $\rho^*_{\leq \epsilon} \in {\rm Int}(P^{2BE}_{\leq \epsilon})$, then there exists $\epsilon_1 < \epsilon$ such that $\rho^*_{\leq \epsilon} \in P^{2BE}_{\leq \epsilon_1}$. As a result, $f(\rho^*_{\leq \epsilon}) \geq f(\rho^*_{\leq \epsilon_1})$. But we also have $f(\rho^*_{\leq \epsilon}) \leq f(\rho^*_{\leq \epsilon_1})$ by set inclusion, such that $W^{2BE}_{\leq \epsilon} = W^{2BE}_{\leq \epsilon_1}$ in contradiction with the strict monotonicity.\\

\noindent As a result $\rho^*_{\leq \epsilon} \in \partial P^{2BE}_{\leq \epsilon}$ and $W_{\epsilon}^{2BE}=W_{\leq \epsilon}^{2BE}$.

\section{Partial bit erasure - product system} \label{app:partialerasureimpth}

\begin{figure}[htbp]
    \label{fig:partialtwobit}
    \centering
    % First subfigure
    \begin{subfigure}[t]{0.45\textwidth}
        \centering
        \includegraphics[width=\textwidth]{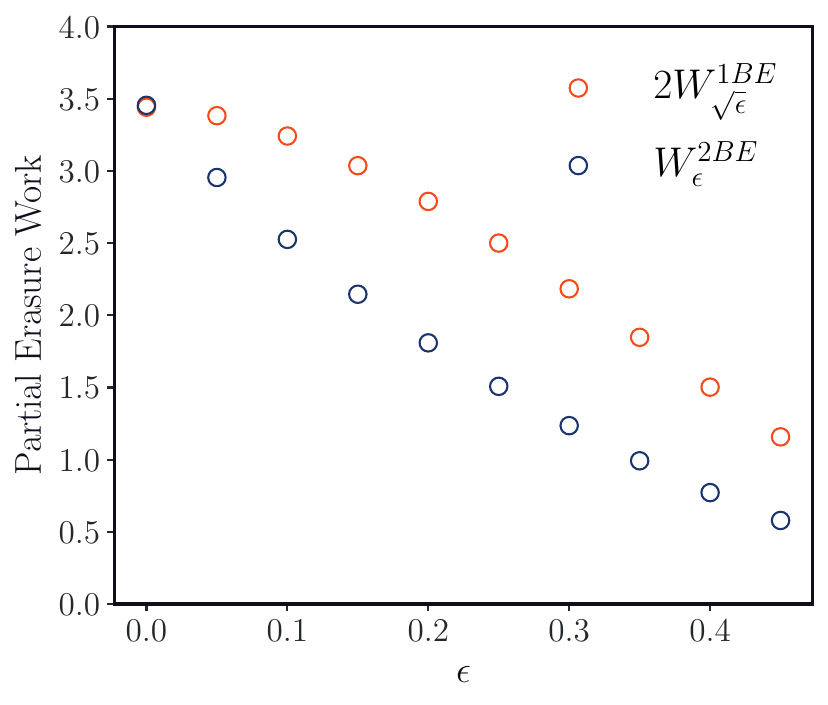}
        \caption{\footnotesize Partial two bit erasure gate : the optimal work (red) is upper bounded by the work associated to two one dimensional gates \eqref{eq:partialupper}.}
        \label{fig:twobitpartialwork}
    \end{subfigure}
    \hfill
    % Second subfigure
    \begin{subfigure}[t]{0.45\textwidth}
        \centering
    \includegraphics[width=\textwidth]{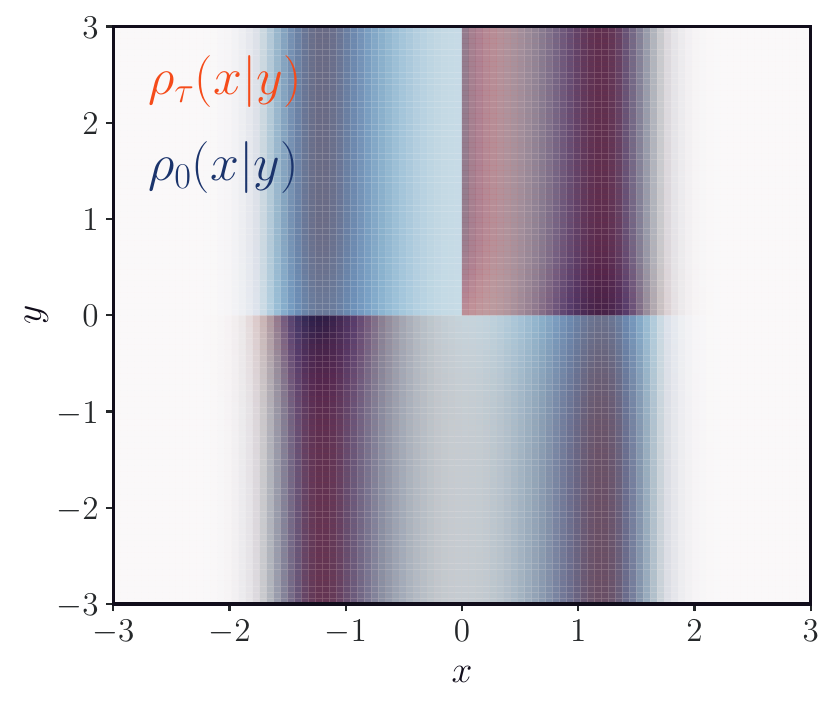}
        \caption{\footnotesize Two dimensional histogram of the optimal conditional target distribution $\rho_\tau(x|y)$. The obvious $y$ dependence at time $t=\tau$ (red) - absent at time $t=0$ (blue) clearly indicates that the target distribution is not a product state}
        \label{fig:twobitpartialcond}
    \end{subfigure}
    \centering
    %\caption{Partial two bit erasure.}
    \label{fig:two_panels}
\end{figure}

In this section we discuss the validity of the factorized erasure theorem \eqref{eq:impth3} in the case of partial two bit erasure

\begin{equation}
W^{\rm 2BE}_\epsilon = \underset{\rho_\tau \in \Rho^{\rm 2BE}_\epsilon}{\inf}\left[T \Dkl(\rho_\tau || \rho_0) + \frac{\was^2(\rho_\tau, \rho_0)}{\tau}\right];~\Rho^{\rm 2BE}_\epsilon = \left\{\rho;\rho(\CC_{00}) = 1-\epsilon \right\}.
\end{equation}
Following the main text, it is clear that for factorized source distribution $\rho_0(\Xb)=\rho_0^x(x)\rho_0^y(y)$ the lower bound
\begin{equation}
\label{eq:partialbound}
\begin{split}
        W^{\rm 2BE}_\epsilon &\geq \underset{\rho_\tau \in \Rho^{\rm 2BE}_\epsilon}{\inf}\left[T \Dkl(\rho^x_\tau || \rho^x_0)  + \tau^{-1}\int_{\RR^2}|x_\tau - x_0|^2\pi_\tau(x_\tau, x_0)\der x_\tau \der x_0 \right]  \\
        &~~~+\underset{\rho_\tau \in \Rho^{\rm 2BE}_\epsilon}{\inf}\left[T \Dkl(\rho^y_\tau || \rho^y_0 ) + \tau^{-1}\int_{\RR^2}|y_\tau - y_0|^2\pi_\tau(y_\tau, y_0)\der y_\tau \der y_0\right]\\
    \end{split}
\end{equation}
holds, where $\rho_\tau^{\{x,y\}}$ denote the target marginals. From the one dimensional partial erasure problem, one is tempted to construct a factorized target distribution $\rho_\tau(\Xb) = \rho^{\sqrt{\epsilon},x}_\tau(x)\rho_\tau^{\sqrt{\epsilon},y}(y)$ where 

\begin{equation}
\rho^{\epsilon,x}_\tau(x) = \underset{\mu \in \Rho^{\rm 1BE}_\epsilon}{\rm arginf}\left[T \Dkl(\mu || \rho^x_0) + \frac{\was^2(\mu, \rho_0^x)}{\tau}\right];~\Rho^{\rm 1BE}_\epsilon = \left\{\rho;\rho(\CC_{1}) = \epsilon\right\},
\end{equation}
such that $\rho_\tau(\CC_{00}) = 1-\epsilon$. In turn, one obtains an immediate upper bound on the two bit work, illustrated on Fig. \ref{fig:twobitpartialwork} :

\begin{equation}
\label{eq:partialupper}
    W^{2BE}_\epsilon(\rho_0^x \rho_0^y) \leq W^{1BE}_{\sqrt{\epsilon}}(\rho_0^x) + W^{1BE}_{\sqrt{\epsilon}}(\rho_0^y).
\end{equation}
However, unlike in the complete erasure case, one does not have

\begin{equation}
\begin{split}
    &\underset{\rho_\tau \in \Rho^{\rm 2BE}_\epsilon}{\inf}\left[T \Dkl(\rho^x_\tau || \rho^x_0)  + \tau^{-1}\int_{\RR^2}|x_\tau - x_0|^2\pi_\tau(x_\tau, x_0)\der x_\tau \der x_0 \right] =\\
    &~~\underset{\rho_\tau^x \in \Rho^{\rm 1BE}_{\sqrt{\epsilon}}}{\inf}\left[T \Dkl(\rho^x_\tau || \rho^x_0)  + \tau^{-1}\int_{\RR^2}|x_\tau - x_0|^2\pi_\tau(x_\tau, x_0)\der x_\tau \der x_0 \right],
    \end{split}
\end{equation}
such that the factorized source-target pair does not saturate \eqref{eq:partialbound}. To reformulate, the optimal target distribution associated to partial two bit erasure is not a product distribution, which is readily seen by computing the associated optimal coupling $\pi_{\rm 2BE}^\epsilon$ and visualizing conditional distributions, as is done on Fig. \ref{fig:twobitpartialcond}.

\section{Dual updates for partial erasure and NAND gate} 
\label{app:dualupdate}

In this section, we explicitly compute the dual updates associated to the partial bit erasure and NAND gate. We recall that the optimal coupling takes the form

\begin{equation}
    \pi(\Xb_\tau, \Xb_0) = e^{\frac{u(\Xb_\tau)}{\epsilon}}K(\Xb_\tau, \Xb_0)e^{\frac{v(\Xb_\tau)}{\epsilon}}
\end{equation}
with $K = e^{-\frac{c}{\epsilon}}$ and $u,v : \RR^2 \to \RR$ the unknown dual functions identified via the iterative scheme

\begin{equation}
    \label{eq:itscheme2}
    \begin{split}
    &u^{\ell +1} = \underset{u}{\rm argmax}\left[-\Tilde{F}_0(-u) - \epsilon \langle e^{\frac{u^\ell}\epsilon},(K*e^{\frac{v^\ell}{\epsilon}})\rangle\right]\\
    &v^{\ell +1} = \underset{v}{\rm argmax}\left[-\Tilde{F}^G_\tau(-v) - \epsilon \langle e^{\frac{v^\ell}\epsilon},(K^T*e^{\frac{u^{\ell+1}}{\epsilon}})\rangle\right],
    \end{split}
\end{equation}
with the Legendre transform of the gate contraint defined as $\Tilde{F}_\tau^G(u) = \underset{p}{\sup}\left[\langle p,u\rangle - F_G(p)\right]$. We hereafter only consider the $v$ update, as the $u$ update is identical to the complete bit erasure gate.

\subsection{Partial bit erasure}

In the partial bit erasure case the gate constraint reads

\begin{equation}
    \begin{split}
        &F_G(p) = T \Dkl(p|| \rho_0) + \iota_{\alpha}(p)\\
        &\iota_{\alpha}(p) = \left\{\begin{split}
            &0\ \text{if}\  p(\CC_{00}) = \alpha\\
            &+\infty\ \text{else}
        \end{split}\right.
    \end{split}
\end{equation}
Since the target set is fixed, we denote the indicator function $\mathbbm{1}(\Xb) \equiv \mathbbm{1}_{\CC_{00}}(\Xb)$ and proceed to compute the Legendre transform by introducing a lagrange multiplier $\lambda$:

\begin{equation}
    \Tilde{F}(u) = \underset{p}{\sup}\left[\langle p,u\rangle - T \Dkl(p||\rho_0) + \lambda (\langle p, \mathbbm{1}\rangle - \alpha\right].
\end{equation}
Recall that the scalar product is defined as
$\sca{f}{g} = \int_{\RR^2}f(\Xb)g(\Xb)\der \Xb$.
We first compute the first variation with respect to $p$:

\begin{equation}
    \begin{split}
        &0 = \frac{\delta}{\delta p}\left[\sca{p}{u} - T \Dkl(p||\rho_0) + \lambda (\sca{p}{\ind}-\alpha)\right]\\
        &0 =\frac{\delta}{\delta p}\left[\sca{p}{u} - T (\sca{p,\log(\frac{p}{\rho_0})} - \sca{p}{1} + \sca{\rho_0}{1}) + \lambda (\sca{p}{\ind}-\alpha)\right]\\
        &0 =\left[u - T \log(\frac{p}{\rho_0})+ \lambda \ind\right]\\
        \implies p = \rho_0 e^{\frac{u}{T} + \lambda \ind}
    &\end{split}.
\end{equation}
Enforcing the Lagrange multiplier constraint then leads to
the arginf of the Legendre transform :

\begin{equation}
    p^*\left[u\right] = \rho_0 e^{\frac{u}{T}}\exp(-\mathbbm{1}\langle[\log(\alpha) - \log(\sca{\rho_0 e^{\frac{u}{T}}}{\ind})\rangle])
\end{equation}
Such that it is clear that $\sca{p^*[u]}{\ind}=\alpha$. For clarity, we rewrite $A[u] = \sca{\rho_0 e^{\frac{u}{T}}}{\ind}$ and inject into the Legendre transform

\begin{equation}
    \begin{split}
        \Tilde{F}(u) &= \sca{u}{p^*[u]} - T \Dkl(p^*[u]||\rho_0)\\
        &=\sca{u}{\rho_0 e^{\frac{u}{T}}e^{-\ind \log(A[u]/\alpha)}} - T \sca{\rho_0 e^{\frac{u}{T}}e^{-\ind \log(A[u]/\alpha)}}{\frac{u}{T} - \ind \log(A[u]/\alpha)-1}\\
        &=T \sca{\rho_0 e^{\frac{u}{T}}e^{-\ind \log(A[u]/\alpha)}}{\ind \log(A[u]/\alpha)+1}\\
        &=T \sca{\rho_0 e^{\frac{u}{T}}}{(1-\ind)} + T(\log(A[u]/\alpha)+1)\frac{\alpha}{A[u]}\sca{\rho_0 e^{\frac{u}{T}}}{\ind}\\
        &=T \sca{\rho_0 e^{\frac{u}{T}}}{(1-\ind)} + T\alpha(\log(A[u]/\alpha)+1)
    \end{split}
\end{equation}
To make the update, we now seek the argmax
\begin{equation}
    \underset{u}{\rm argmax}\left[-\Tilde{F}(-u) - \epsilon \sca{e^{\frac{u}{\epsilon}}}{(K*e^{\frac{v}{\epsilon}})}\right].
\end{equation}
Taking a functionnal derivative yields
\begin{equation}
\label{app:eq-upartial}
    \begin{split}
        &0 = \frac{\delta}{\delta u}\left[-\Tilde{F}(-u) - \epsilon \sca{e^{\frac{u}{\epsilon}}}{(K*e^{\frac{v}{\epsilon}})}\right]\\
        &0 = \frac{\delta}{\delta u}\left[-T \sca{\rho_0 e^{\frac{-u}{T}}}{(1-\ind)} - T\alpha(\log(A[-u]/\alpha)+1) - \epsilon \sca{e^{\frac{u}{\epsilon}}}{(K*e^{\frac{v}{\epsilon}})}\right]\\
        &0 = \rho_0 e^{\frac{-u}{T}}(1-\ind) + \alpha\frac{1}{A[-u]}\rho_0 e^{\frac{-u}{T}}\ind -  e^{\frac{u}{\epsilon}}(K*e^{\frac{v}{\epsilon}})\\
        &0 = \rho_0 e^{\frac{-u}{T}}\left((1-\ind) + \frac{\alpha}{A[-u]}\ind\right) -  e^{\frac{u}{\epsilon}}(K*e^{\frac{v}{\epsilon}})\\
        &\rho_0 e^{\frac{-u(T + \epsilon)}{T \epsilon}}\left((1-\ind) + \frac{\alpha}{A[-u]}\ind\right) =  (K*e^{\frac{v}{\epsilon}})\\
    \end{split}
\end{equation}
We first determine $A[-u]$ self consistently from the previous expression :

\begin{equation}
\label{app:eq-normupdate}
    \begin{split}
        &\ind \rho_0 e^{-u \frac{T + \epsilon}{T \epsilon}} = \frac{A[-u]}{\alpha}(K*e^{\frac{v}{\epsilon}})\ind\\
        &\ind \rho_0 e^{-\frac{u}{T}} = \left(\frac{A[-u]}{\alpha}\right)^{\frac{\epsilon}{T+\epsilon}}(K*e^{\frac{v}{\epsilon}})^{\frac{\epsilon}{T+\epsilon}}\rho_0^{\frac{T}{T+\epsilon}}\ind\\
        &A[-u]^{\frac{T}{T+\epsilon}} = \left(\frac{1}{\alpha}\right)^{\frac{\epsilon}{T+\epsilon}}\sca{(K*e^{\frac{v}{\epsilon}})^{\frac{\epsilon}{T+\epsilon}}\rho_0^{\frac{T}{T+\epsilon}}}{\ind}\\
        &\log(A[-u]) = \frac{T+\epsilon}{T} \log\left(\sca{\left(\frac{K*e^{\frac{v}{\epsilon}}}{\alpha}\right)^{\frac{\epsilon}{T+\epsilon}}\rho_0^{\frac{T}{T+\epsilon}}}{\ind}\right).
    \end{split}
\end{equation}
Because $A[-u]$ only depends on the data $\rho_0$, distance matrix via $K$ and prior dual vector $v$ it can be readily computed from the previous expression. In turn, one obtains two different update for $u$ from \eqref{app:eq-upartial}:

\begin{equation}
\label{app:eq-fuupdate}
    \begin{split}
        &\ind(\Xb) = 0 \implies u = \frac{T \epsilon}{T + \epsilon}\left[\log(\rho_0) - \log(K*e^{\frac{v}{\epsilon}})\right]\\
        &\ind(\Xb) = 1 \implies u = \frac{T \epsilon}{T + \epsilon}\left[\log(\rho_0) - \log(K*e^{\frac{v}{\epsilon}}) + \log(\alpha) - \log(A[-u])\right]\\.
    \end{split}
\end{equation}
Taken together, the updates \eqref{app:eq-normupdate} and \eqref{app:eq-fuupdate} yield the dual variable $u$.

\subsection{NAND gate}

The NAND gate combines aspect from both the partial erasure and complete erasure setups. Specifically the associated constraint is written as

\begin{equation}
    \begin{split}
        &F_G(p) = T \Dkl(p|| \rho_0) + \iota_{\rm NAND}(p)\\
        &\iota_{\rm NAND}(p) = \left\{\begin{split}
            &0\ \text{if}\  p(\CC_{11}) = 0.25~{\rm and}~p(\CC_{01}\bigcup\CC_{10}) = 0\\
            &+\infty\ \text{else}.
        \end{split}\right.
    \end{split}
\end{equation}
To reformulate, the NAND gate seeks to erase the mass on one of the diagonals and balance it in a 0.25/0.75 ratio on the remaining two quadrants. Combining the computations from the partial erasure and complete erasure gate, the $u$ and $v$ update are easily shown to take the compact following form

\begin{equation}
    \begin{split}
        &v^{\ell + 1} = \epsilon\left[\log(\rho_0) - \log(K*e^{\frac{u^{\ell}}{\epsilon}})\right]\\
        &u^{\ell + 1} = \frac{T}{T+\epsilon}\left[\log(\rho_0 \ind_{C_{00}\bigcup C_{11}}) - \log(K*e^{\frac{v^{\ell + 1}}{\epsilon}}) + \ind_{C_{11}}\left[\log(0.25) - A\right] \right]\\
        &A = \frac{T+\epsilon}{T} \log\left(\sca{\left( 4 K*e^{\frac{v}{\epsilon}}\right)^{\frac{\epsilon}{T+\epsilon}}\rho_0^{\frac{T}{T+\epsilon}}}{\ind_{C_{11}}}\right).
    \end{split}
\end{equation}
where the convention is again $\log(0) = -\infty$.

\section{Scaling forms, LogSumExp tricks}

In this section we briefly discuss the classical 'scaling' regularized unbalanced optimal transport implementation, and some of the numerical choices we make to compute regularized coupling.

\subsection{Scaling forms.} 

In their most widespread versions \cite{Chizat2018}, the iterative schemes described above are usually carried out directly on the scaling variables $(a^{\ell}, b^{\ell}) = (e^{\frac{u^{\ell}}{\epsilon}},e^{\frac{v^{\ell}}{\epsilon}})$, for which the updates are given by

\begin{equation}
    \begin{split}
        &a^{\ell+1} = \underset{a}{\rm argmin}\left[F_0(a)+\epsilon \Dkl(a|| K*b^{\ell})\right] / (K*b^{\ell})\\
        &b^{\ell+1} = \underset{b}{\rm argmin}\left[F_\tau(b)+\epsilon \Dkl(b|| K*a^{\ell + 1})\right] / (K*a^{\ell +1})\\
    \end{split}
\end{equation}
where the division is carried out entry-wise. The term scaling form is used because for many functions $F$  of interest, the update is fully explicit in terms of matrix multiplications. For instance, the fixed marginal update associated to the original OT problem $F_0(a) = \iota_{a = \rho_0}$ simply yields the update

\begin{equation}
    a^{\ell + 1} = \rho_0 / (K*b^{\ell}).
\end{equation}
Under the scaling form, the gate updates take a clear physical interpretation. In the NAND case, the $b^{\ell+1}$ update can be decomposed into 4 steps:

\begin{itemize}
    \item Computing the explicit scaling argmin without the mass constraints, first set 
    \begin{equation}
        b^{\ell+1} = \rho_0 ^{\frac{T}{T+\epsilon}}(K*a^{\ell+1}) ^{\frac{\epsilon}{T+\epsilon}}
    \end{equation}
    \item For all $\Xb \in \CC_{01}\bigcup C_{10}$ set $b^{\ell+1}(\Xb)=0$.
    \item Normalize $b^{\ell}$ over the set $\CC_{11}$ : for $\Xb \in \CC_{11}$ set  $b^{\ell+1}(\Xb) \leftarrow b^{\ell+1}(\Xb) / \sca{b^{\ell+1}}{\ind_{\CC_{11}}}\times 0.25$
    \item Divide entry-wise $b^{\ell+1} \leftarrow b^{\ell+1} / (K*a^{\ell + 1})$
\end{itemize}
Albeit much simpler and computationally faster than the dual update, the scaling update suffers from numerical instabilities as soon as $\epsilon$ becomes small. In the stochastic thermodynamic framework we are considering, reaching small $\epsilon$ gate accuracy is paramount, hence the emphasis on dual updates.

\subsection{LogSumExp trick} Computing terms of the form $\log(K * e^{\frac{v}{\epsilon}})(\Yb) = \log(\int  e^{\frac{c(\Yb, \Xb)}{\epsilon}}e^{\frac{v(\Xb)}{\epsilon}}\der \Xb)$ can lead to numerical instabilities due to exploding values of $(c(\Yb, \Xb) + v(\Xb))/\epsilon$. A convenient numerical trick dubbed the LogSumExp trick is to first compute the max $m(\Yb) = \underset{\Xb}{\rm max}\frac{c(\Yb, \Xb) + v(\Xb)}{\epsilon}$ and reexpress the previous integral as

\begin{equation}
    \begin{split}
        \log(K*e^{\frac{v}{\epsilon}})(\Yb) = m(\Yb) + \log(\int e^{\frac{c(\Yb, \Xb)}{\epsilon}}e^{\frac{v(\Xb)}{\epsilon}}e^{-m(\Yb)}\der \Xb)
    \end{split},
\end{equation}
such that all exponentiated values are negative. We systematically use the LogSumExp in the dual updates, and use a nested version  twice to compute the normalization constant $A[u]$

\section{Numerical details}

In this section we provide details on various hyperparameters and numerical procedures used to produce figures throughout the main text.

\paragraph{Static coupling}

In practice all computations are carried out on a discrete grid. In turn, functions $p(\Xb)$ become vectors $(p_i)_{i\in [0,N]}$. For discrete distributions,  we take the normalization convention $1_N^T. \rho_0  = 1$, where  $1_N$ is a vector of size $N$ containing all ones.\\

\noindent The two dimensional source and target distribution associated to the two dimensional logical gates are discretized on grids of $80 \times 80$ points.
The unbalanced optimal transport iterative scheme is ran until the convergence criterion $||\pi^{\ell+1} - \pi^{\ell}||_\infty<10^{-5}$ is met or the maximum number of steps $n_{\rm max}=10^4$ is reached.
Marginals are plotted by computing either the left or right unit vector product $\pi.1_{80\times 80}$, $1^T_{80\times 80}\pi$.

\paragraph{Flow matching}

To identify the noiseless optimal flow field $v_\theta$ we prepare data pairs from the optimal coupling.
Specifically we construct grid based data points $(x^i_0, x^i_1)$ by multinomial sampling of $\pi$. We artificially increase sample diversity by construct a continuous dataset $\{(x^i_0 + \epsilon^i, x_1^i + \epsilon^i_1)\}_{1\leq i \leq n_{\rm data}}$ with $n_{\rm data} = 3.10^4$ and $\epsilon\sim \mathcal{N}(0,0.01)$. \\

\noindent The flow model $v_\theta$ is trained using the flow matching objective \eqref{eq:otfm1}.
We use a fully connected MLP with $n_{\rm hidden layers} = 1$ hidden layers, $n_{\rm embed} = 256$ and ReLu activation function.
The MLP is optimized for $n_{\rm epochs}=2000$ using a batch size of $n_{\rm batch}=64$ with Adam and learning rate ${\rm lr}=10^{-3}$.

\paragraph{Score estimation}

For the score estimation needed for Fig \ref{fig:nanddynamical}, we use scipy's RegularGridInterpolator over smoothed empirical histograms.\\

\noindent Specifically, we populate $n_{\rm hist} = 100$ two-dimensional histograms containing $n_{\rm bins}= 50\times 50$ by  propagating source samples $\{x_0\}$ up to time $t^k = k \tau / n_{\rm hist}$ using the learned flow field $v_\theta$.
Each histogram is smoothed using a gaussian filter with varying smoothing parameter (between 0.1 and 1) depending on the considered $\tau$.
Finally, we represent the score as a 2 + 1 space time linear RegularGridInterpolator interpolating the data $\nabla \log \rho_{\rm smooth}(x_i, y_j,t_k)$.

\bibliographystyle{unsrtnat} % We choose the "plain" reference style
\bibliography{refs, references, refs_2} % Entries are in the refs.bib file

\end{document}